\documentclass[11pt]{article}
\usepackage{pictex}
\usepackage{graphicx}
\usepackage[numbers,sort&compress]{natbib}
\usepackage[dvipsnames]{xcolor}
\usepackage[letterpaper,margin=1in]{geometry}
\begin{document}
\title{Using Genetic Data to Build Intuition about Population History}
\author{Alan R. Rogers}
\date{\today}
\maketitle

\begin{abstract}
Genetic data are now routinely used to study the history of population
size, subdivision, and gene flow. A variety of formal statistical
methods is available for testing hypotheses and fitting models to
data. Yet it is often unclear which hypotheses are worth testing,
which models worth fitting. There is a need for less formal methods
that can be used in exploratory analysis of genetic data.  One
approach to this problem uses \emph{nucleotide site patterns}, which
provide a simple summary of the pattern in genetic data. This article
shows how to use them in exploratory data analysis.
\end{abstract}

\section{Nucleotide site patterns}
\label{sec.sitepat}
Nucleotide site patterns are often used in computer programs designed
to infer population history from genetic data
\citep{Rogers:PNA-114-9859, Rogers:BMC-20-526, Rogers:SA-6-eaay5483,
  Rogers:bioRxiv-2021.01.23.427922}. This article however is not
concerned with computer programs. The goal here is to see what can be
learned about population history simply by looking at the data. This
introductory section will introduce site patterns and explain how they
relate to history.

\begin{figure}
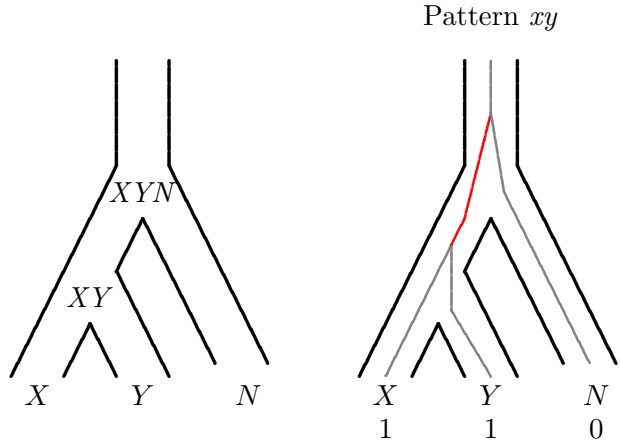

{\centering\input{fignotation}\hspace{1cm}\input{figcongruent}\\}
\caption{Population trees with notation (left) and an embedded gene
  genealogy (right). Population \textit{XY} is ancestral to $X$ and
  $Y$, and \textit{XYN} is ancestral to \textit{XY} and $N$.  At the
  bottom of the right panel, ``0'' indicates the ancestral allele and
  ``1'' the derived (or mutant) allele. A mutation on the solid red
  segment would generate the \textit{xy} site pattern. In this
  example, the closest relatives ($X$ and $Y$) uniquely share the
  derived allele.}
\label{fig.congruent}
\end{figure}

The left panel of Fig.~\ref{fig.congruent} shows the relationship
between three populations. I use capital letters to refer to
populations: $X$ is African, $Y$ is Eurasian, and $N$ is
Neanderthal. Later on we will encounter $D$, which refers to
Denisovans, an archaic population related to Neanderthals
\citep{Reich:N-468-1053}. Combinations of letters refer to ancestral
populations: \textit{XY} is ancestral to $X$ and $Y$, \textit{XYN} is
ancestral to \textit{XY} and $N$, and so on.  The right panel of
Fig.~\ref{fig.congruent} shows the gene genealogy of a tiny sample
consisting of three random nucleotides, one from $X$, one from $Y$,
and one from $N$. All three are drawn from the same position within
the genome. In practice we work with larger samples, involving
multiple individuals and millions of nucleotide sites, but this tiny
sample makes the principles easier to explain. I will call such
samples ``haploid,'' because the sample from each population consists
of a single nucleotide.

Haploid samples also provide a real advantage. When we trace larger
samples backwards into the past, it occasionally happens that two
lineages coalesce at their common ancestor to form a single ancestral
lineage. This happens faster in small populations than in large ones,
so the process is sensitive to population size. With a haploid sample,
no coalescent events are possible until we reach an ancestral
population that contains the ancestors of two or more of our modern
samples. Consequently, recent changes in population size have no
effect, and we need not incorporate them into our model. This makes it
easier to study the distant past.

Figure~\ref{fig.congruent} shows the gene genealogy of one
hypothetical haploid sample, which is represented by the red and the
gray lines. For this sample, the lineages from $X$ and $Y$ coalesce in
population $XY$. Suppose that a mutation occurred somewhere on the red
segment of the genealogy. A mutation generates a new allelic state,
which we will call \emph{derived} or \emph{mutant} to distinguish it
from the \emph{ancestral} state that came before. If a mutation
occurred on the red segment, we would observe the derived allele in
the samples from $X$ and $Y$ and the ancestral allele in that from
$N$.  I will call this outcome the ``\textit{xy} site pattern,'' using
lower case to distinguish site patterns from populations.

\begin{figure}
  {\centering\input{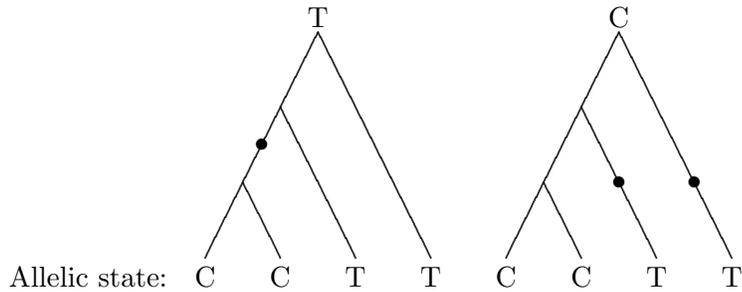}\\}
  \caption{Calling ancestral and derived alleles. The solid lines
    represent a gene genealogy, with observed allelic states at the
    bottom. The letter at the top of each graph is an hypothesis about
    which allele is ancestral. The right-most sample in each graph is
    from an ``outgroup''---a population that is distantly related to
    others. If $C$ were ancestral, it would take two mutations (shown
    as bold dots) to produce the pattern in the data. Only one
    mutation is needed if $T$ is ancestral. Because mutations are
    rare, a model that requires only one mutation is a better bet than
    one requiring two. We therefore infer that $T$ is ancestral and
    $C$ is derived.}
  \label{fig.ancder}
\end{figure}  

Before we can recognize site patterns in the data, we must first
determine which allele is ancestral and which is
derived. Figure~\ref{fig.ancder} shows how this is done.

\section{Data}
\label{sec.data}
\begin{figure}
{\centering \includegraphics[width=0.5\linewidth]{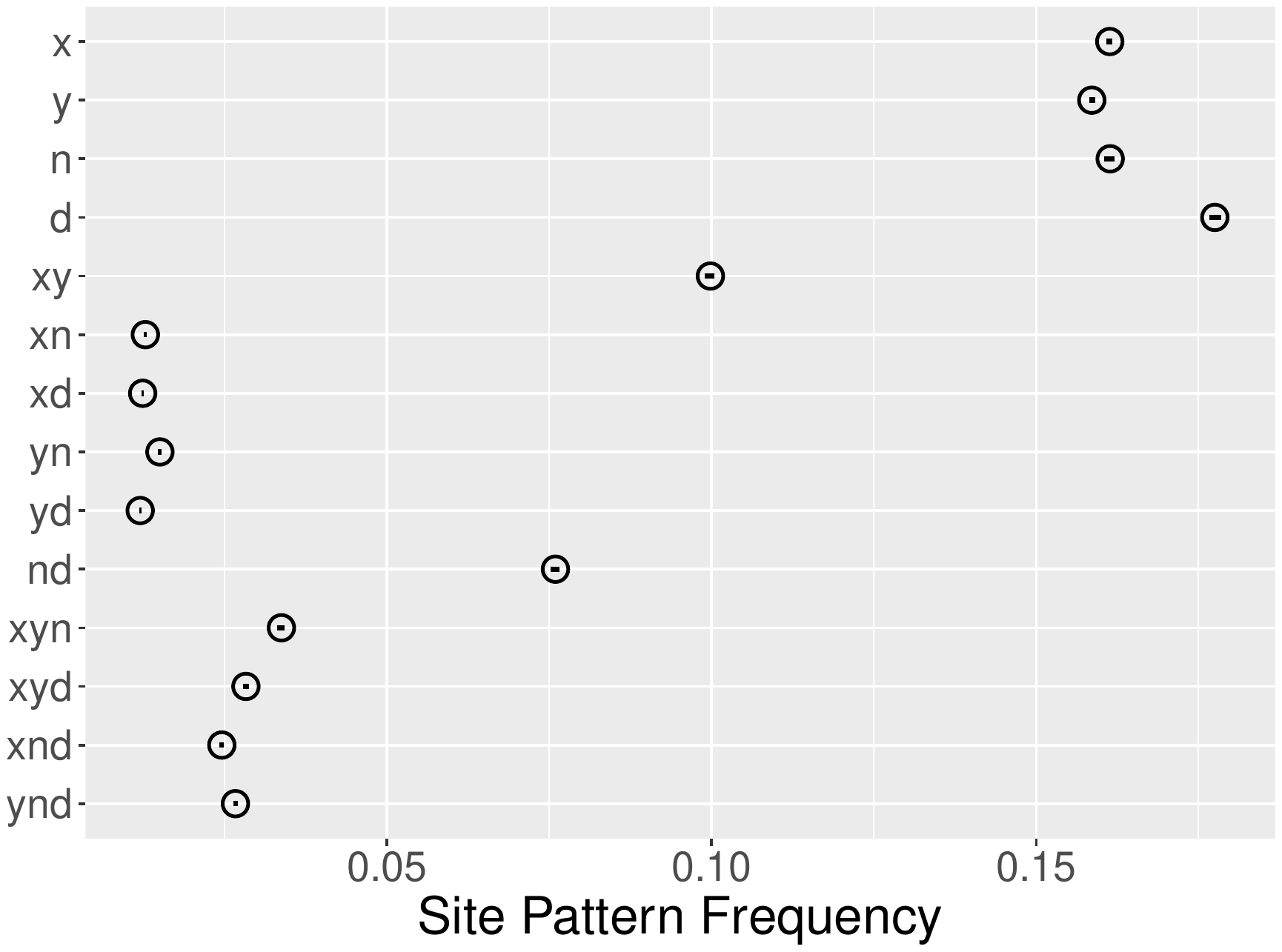}\\}
\caption{Observed site pattern frequencies. Horizontal axis shows the
  relative frequency of each site pattern in random samples consisting
  of a single haploid genome from each of four populations $X$
  (Africa), $Y$ (Europe), $N$ (Neanderthal), and $D$ (Denisovan).
  Horizontal lines (which look like dots) are 95\% confidence
  intervals estimated by a moving-blocks bootstrap
  \citep{Liu:ELB-92-225}. Data: Simons Genome Diversity Project
  \citep{Mallick:N-538-201} and Max Planck Institute for Evolutionary
  Anthropology \citep{Prufer:S-10.1126-science.aao1887}. After
  \citet[Fig.~S2]{Rogers:SA-6-eaay5483}.}
\label{fig.xynd-frq}
\end{figure}

Figure~\ref{fig.xynd-frq} shows site pattern frequencies of two modern
human populations ($X$, Africa, and $Y$, Europe), and two archaic
ones ($N$, Neanderthal, and $D$, Denisovan). In these data, the
horizontal axis shows the relative frequency of each site pattern
among the nucleotide sites in the genome, after excluding sites that
fail quality control criteria, or for which the ancestral and derived
alleles cannot be identified. The frequencies are plotted as open
circles. Within each circle, the thing that looks like a dot is
actually a horizontal line representing the 95\% confidence
interval. These intervals are very narrow, because these frequencies
are estimated with high precision. This precision is possible because
there is so much data in the nuclear genome.

Some of the pattern in these data is obvious. There are a lot of
singletons ($x$, $y$, $n$, and $d$), and two of the doubletons
(\textit{xy} and \textit{nd}) are more common than the others. Of the
two common doubletons, \textit{xy} is more common than
\textit{nd}. Other parts of the pattern are subtle: the less-common
doubletons are similar in frequency, but \textit{yn} is a little more
common, and this difference is larger than the (barely-perceptible)
confidence intervals.

To interpret these patterns, the trick is to reason as we did in
Fig.~\ref{fig.congruent} above. There, a mutation anywhere on the red
segment of the gene genealogy would generate the \textit{xy} site
pattern. Anything that tended to lengthen this segment would inflate
the frequency of this site pattern, because mutations are more likely
to occur on long segments than on short ones. In the section that
follows, we reason from population history to gene genealogies in
order to understand the frequencies of site patterns.

\section{Relating population history to site pattern frequencies}
\subsection{When the gene genealogy agrees with the population tree}
\begin{figure}
  {\centering\input{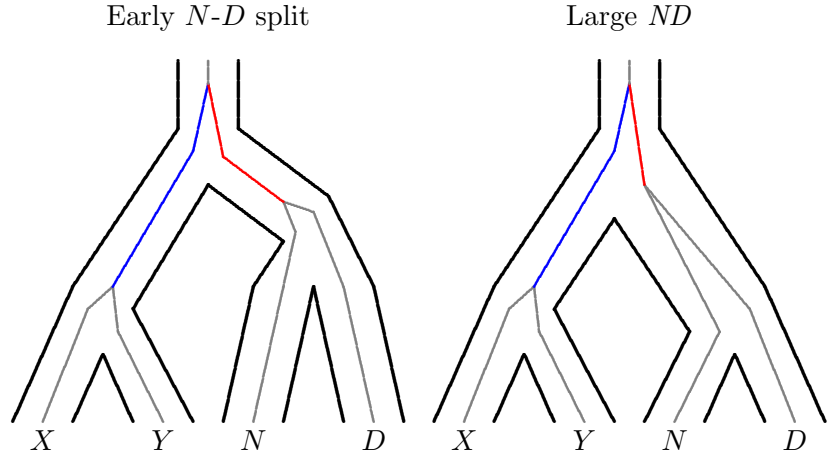}\\}
  \caption{Gene genealogies that match the population tree. $D$ is the
    Denisovan population. Other symbols are as in
    Fig.~\ref{fig.congruent}. The left and right panels illustrate
    different assumptions about the separation times of $X$ and $Y$
    and of $N$ and $D$, and about the effective sizes of populations
    \textit{XY} (ancestral to $X$ and $Y$) and \textit{ND} (ancestral
    to $N$ and $D$). The left panel assumes that the two separation
    times are unequal but the effective sizes of the two ancestral
    populations are the same. The right panel assumes the two
    separation times are the same but that the effective size of
    population \textit{ND} is larger than that of \textit{XY}.}
  \label{fig.2fork}
\end{figure}

It is useful to start with a model that lacks gene flow. Such models
often shed light on large-scale patterns, even if they miss the
subtleties. Figure~\ref{fig.2fork} shows two versions of such a model,
each of which assumes that the two modern human populations ($X$ and
$Y$) are closely related, and so are the two archaic ones ($N$ and
$D$). Because there is no gene flow in these models, the network of
populations has the shape of a tree. Within each population tree, I
have drawn a gene genealogy, and in each genealogy the branching order
is the same as that of the population tree. As we will see below,
other branching orders also occur. However, this one is likely to be
common, and if so it will explain the major features in the data.

In both sides of Fig.~\ref{fig.2fork}, a mutation on the blue segment
would generate site pattern \textit{xy}, whereas a mutation on the red
segment would generate \textit{nd}. These are the two common
doubletons in Fig.~\ref{fig.xynd-frq}. It makes sense that these
patterns should be common, because $X$ (Africa) and $Y$ (Europe) are
close relatives, as are $N$ (Neanderthal) and $D$ (Denisovan). Close
relatives tend to share ancestors, and mutations in these shared
ancestors inflate the frequencies of \textit{xy} and \textit{nd}. This
is the verbal intuition behind the graphs in Fig.~\ref{fig.2fork}. The
data imply that the two modern populations are close relatives as are
the two archaic ones.

But why is \textit{xy} more common than \textit{nd}?  In
Fig.~\ref{fig.2fork}, site pattern \textit{xy} will arise only if a
mutation occurs on the blue branch. This is more likely to happen if
the blue branch is long than if it is short. Thus, the excess of
\textit{xy} over \textit{nd} implies that, on average across the
genome, the blue branch is longer than the red one. Why should these
branches tend to differ in length?

The left and right panels of Fig.~\ref{fig.2fork} illustrate two
possible reasons. In the left panel, the separation between $X$ and
$Y$ is more recent than that between $N$ and $D$. For this reason, $X$
and $Y$ share a longer history of common ancestry than do $N$ and
$D$. Consequently, the blue branch is longer than the red one, and
site pattern \textit{xy} is more common than \textit{nd}, just as we
see in the data.

This is reassuring and probably also correct, but I have smuggled in a
hidden assumption. As we trace the lineages from $X$ and from $Y$
backwards in time, they eventually arrive in population
\textit{XY}. Similarly, the lineages from $N$ and $D$ arrive in
population \textit{ND}, which is ancestral to $N$ and $D$. In either
case, as we trace the two lineages farther back in time, they
eventually coalesce into a single parental lineage. In the previous
paragraph, I assumed that this process takes the same length of time
within populations \textit{XY} and \textit{ND}. This is not
necessarily so.

Gene genealogies tend to be deeper in large populations than in small
ones. Why? Because two random individuals are more likely to be close
relatives in a small population. Close relatives are connected by
short genealogies and distant relatives by deep ones. In the right
panel of Fig.~\ref{fig.2fork}, population \textit{ND} is much larger
than \textit{XY}, and it takes a long time for the $N$ and $D$
lineages to coalesce after they arrive in \textit{ND}. In other words,
they are connected by a deep genealogy. For this reason, the red
branch is much shorter than the blue one. This is a second hypothesis
that might explain the excess of \textit{xy} over \textit{nd} in the
data.

It is not easy to choose between these alternatives, and this
difficulty plagues formal statistical methods as well as the
exploratory ones discussed here. In spite of this difficulty, we do
have estimates, and these indicate that Neanderthals and Denisovans
separated much earlier than did Africans and Europeans
\citep{Reich:N-468-1053}. Furthermore, my colleagues and I have
estimated that the population ancestral to Neanderthals and Denisovans
was very small \citep{Rogers:PNA-114-9859,
  Rogers:SA-6-eaay5483}. Thus, neither of the assumptions made in the
right panel of Fig.~\ref{fig.2fork} is likely to be correct. The
excess of \textit{xy} over \textit{nd} probably reflects a recent
separation of $X$ and $Y$ rather than large size in population
\textit{ND}.

\subsection{When the gene genealogy disagrees}
\label{sec.linsort}
Having explained several large-scale features in the data, let us
focus now on a finer scale. It will be useful for a little longer to
ignore the effect of gene flow. This will provide a base line against
which effects of gene flow will become visible.

\begin{figure*}
{\centering\input{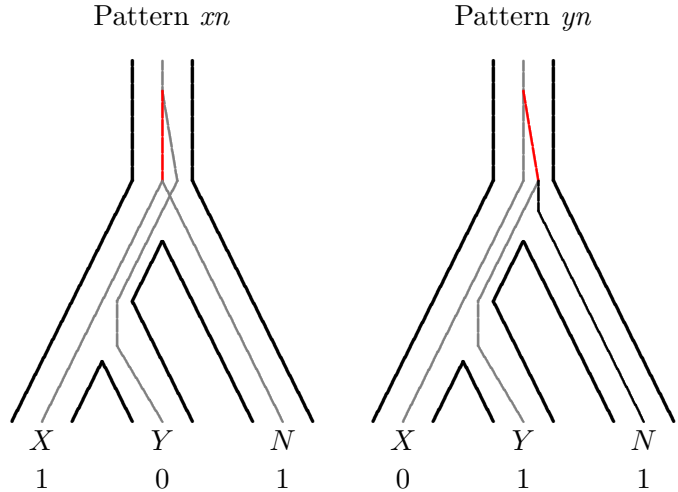}\\}
\caption{Incomplete lineage sorting: distant relatives may share
  derived allele. Absent gene flow, these site patterns should be
  equally frequent.}
\label{fig.linsort}
\end{figure*}

In this section, we concentrate on counterintuitive sites
patterns---ones in which the derived allele is shared not by samples
from closely-related populations, but by distant relatives.  Consider
for example the two histories shown in Fig.~\ref{fig.linsort}. In
these examples, the lineages sampled in $X$ and $Y$ happen---just by
chance---not to coalesce in population $XY$. They persist (going
backwards in time) as distinct lineages throughout this segment of
population history until they reach population \textit{XYN} and are
joined by the lineage sampled from $N$. At this point, we have three
lineages within the larger population. I will assume that this
population is not geographically subdivided. In this case, the three
lineages are equally likely to coalesce in any order: either
$((X,Y),N)$, $((X,N), Y)$, or $(X, (Y,N))$. The two counterintuitive
orders, which are illustrated in Fig.~\ref{fig.linsort}, generate site
patterns \textit{xn} and \textit{yn}. The process that leads to these
counterintuitive patterns is called \emph{incomplete lineage sorting}
\citep{Pamilo:MBE-5-568}.

Incomplete lineage sorting generates site patterns \textit{xy},
\textit{xn}, and \textit{yn} with equal frequency. However,
\textit{xy} is also generated by coalescent events within population
\textit{XY}, as illustrated in Fig.~\ref{fig.congruent}. Thus, in the
absence of admixture, we expect \textit{xn} and \textit{yn} to be
equally common, but neither of these should be as common as
\textit{xy}.

Returning now to the data in Fig.~\ref{fig.xynd-frq}, note that
\textit{xd} is about equal in frequency to \textit{yd}. These are
counterintuitive site patterns in which the derived allele is shared
by samples from distantly-related populations. If these site patterns
were produced by incomplete lineage sorting, they should be equal in
frequency, as indeed they seem to be.

On the other hand, the frequencies of \textit{yn} and \textit{xn} are
not equal. Although the difference is not large in absolute terms, it
is (as noted above) much larger than the confidence intervals. To
understand what this might mean, we need to discuss the effect of gene
flow between populations.

\subsection{The effect of gene flow}
\label{sec.mix}
\begin{figure}
{\centering\input{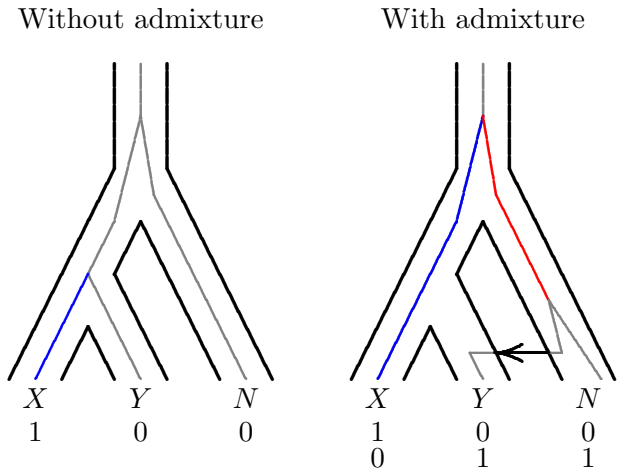}\\}
\caption{Neanderthal admixture into $Y$ inflates the \emph{x} and
  \emph{yn} site patterns. Blue segment generates the \textit{x} site
  pattern, red branch the \textit{yn} site pattern.}
\label{fig.mix}
\end{figure}

Figure~\ref{fig.mix} shows the result of gene flow from the
Neanderthal population ($N$) into the Eurasian one ($Y$). This form of
gene flow inflates the frequencies of the \textit{x} and \textit{yn}
site patterns. In the figure, the blue branch is longer on the right
than on the left, and this increases the frequency of $x$. The red
branch on the right provides a mechanism (in addition to incomplete
lineage sorting) that can generate \textit{yn}. Both patterns can
arise without gene flow, but they are more common when Neanderthals
contribute DNA to Europeans.

These effects are not likely to be large. If only a small fraction of
the lineages in $Y$ derive via gene flow from $N$, then the histories
of most nucleotide sites will be as described in
Figs.~\ref{fig.congruent}, \ref{fig.2fork}, and~\ref{fig.linsort}. In
a complete genome from $Y$, only a small fraction of the nucleotide
sites will derive from $N$. Consequently, the frequencies of site
patterns \textit{yn} and \textit{x} will be inflated only slightly.

This is exactly the pattern seen in Fig.~\ref{fig.xynd-frq}:
\textit{yn} is more common than \textit{xn}, and $x$ is more common
than $y$. Although neither effect is large, both are larger than the
confidence intervals. This pattern is the signature of Neanderthal
gene flow into Europeans \citep{Green:S-328-710}.

Finally, note that the $d$ site pattern is substantially more common
than the other singletons. This suggests that the Denisovan fossil is
younger than the Altai Neanderthal. However, a detailed analysis of
this hypothesis \citep{Rogers:PNA-114-E10258} led to the absurd
conclusion that the Denisovan fossil was only 4000 years old---a date
that is tens of thousands of years too young. Something was clearly
missing from our model. Several authors had suggested that Denisovans
received gene flow from a ``superarchaic'' population, which was
distantly related to all other humans, archaic as well as modern
\citep{Waddell:arXiv-1112.6424, Waddell:arXiv-1312.7749,
  Prufer:N-505-43,Prufer:S-10.1126-science.aao1887,
  Kuhlwilm:N-10.1038-nature16544}. If this hypothesis were correct,
what effect might this have on site pattern frequencies?

\begin{figure}
  {\centering\input{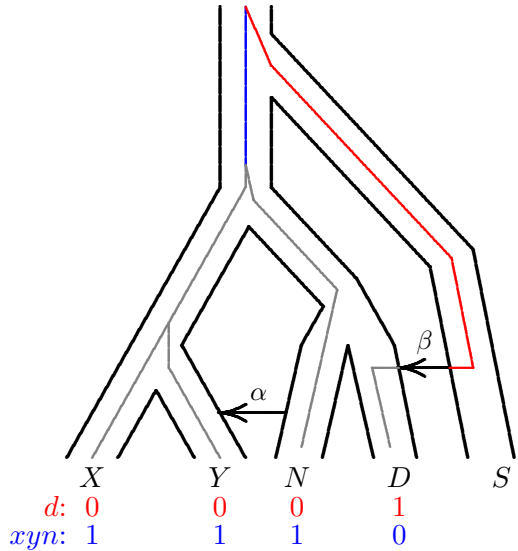}\\}
  \caption{A population network including two episodes of gene flow,
    with an embedded gene genealogy. Upper case letters ($X$, $Y$,
    $N$, $D$, and $S$) represent populations (Africa, Europe, Altai
    Neanderthal, Denisovan, and superarchaic). Greek letters label
    episodes of gene flow. $d$ and $xyn$ illustrate two nucleotide
    site patterns, in which 0 and 1 represent the ancestral and
    derived alleles. A mutation on the red branch would generate site
    pattern $d$. One on the blue branch would generate $xyn$. After
    \citet[Fig.~1]{Rogers:SA-6-eaay5483}.}
  \label{fig.hyparch}
\end{figure}

Figure~\ref{fig.hyparch} provides an answer. Mutations on the red
branch in this figure would inflate the frequency of site pattern
$d$. Mutations on the blue branch would inflate \textit{xyn}. Refer
back to Fig.~\ref{fig.xynd-frq} and you will see that not only is $d$
more common than the other singletons, \textit{xyn} is more common
than the other tripletons. This is the signature of superarchaic
admixture into Denisovans \citep{Rogers:SA-6-eaay5483}.

\section{Conclusions}
This article has tested no hypotheses. Neither has it used any formal
method to fit models to data. Yet we have learned a lot simply by
looking at the data---by using it to build intuition about what models
to fit and what hypotheses to test. Our exploratory analysis suggests
that (a)~modern Europeans and Africans are closely related, (b)~so
were Neanderthals and Denisovans, (c)~the separation between Europeans
and Africans was more recent than that between Neanderthals and
Denisovans, (d)~Neanderthals contributed DNA to Europeans, and
(d)~superarchaics contributed DNA to Denisovans. None of these
findings are new, but it is useful to see that they can all be
inferred simply by looking at the data.

\section*{Acknowledgements}
I am grateful for comments from Elizabeth Cashdan, Touhid Islam, Jan
Ko{\v c}i, and Daniel Tabin. This work was supported by NSF grant
BCS~1945782.

\bibliographystyle{plainnat}
\bibliography{speda}
\end{document}